\documentclass[prb,amsmath,amssymb,twocolumn]{revtex4-2}
\usepackage{graphicx}
\usepackage{subfigure}
\usepackage{adjustbox}
\usepackage{bm}
\usepackage{color}
\usepackage{braket}
\usepackage{standalone}
\usepackage{multirow}
\usepackage{tikz}
\usepackage{mathrsfs}
\usepackage{dsfont}
\usepackage[colorlinks,bookmarks=true,citecolor=blue,linkcolor=blue,urlcolor=blue]{hyperref}
\usepackage{cleveref}
\usepackage{comment}
\usepackage{float}

\begin{document}
\title{Anomalous Skin Effects in Disordered Systems with a Single non-Hermitian Impurity}
\author{Paolo Molignini, Oscar Arandes  and Emil J. Bergholtz}
\affiliation{Department of Physics, Stockholm University, AlbaNova University Center, 10691 Stockholm, Sweden}
\date{\today}

\begin{abstract}
We explore anomalous skin effects at non-Hermitian impurities by studying their interplay with potential disorder and by exactly solving a minimal lattice model. 
A striking feature of the solvable single-impurity model is that the presence of anisotropic hopping terms can induce a { scale-free} accumulation of all eigenstates {\it opposite} to the bulk hopping direction, although the non-monotonic behavior is fine-tuned and further increasing such hopping weakens and eventually reverses the effect. 
The interplay with bulk potential disorder, however, qualitatively enriches this phenomenology leading to a robust non-monotonic localization behavior as directional hopping strengths are tuned. 
 Non-monotonicity persists even in the limit of an entirely Hermitian bulk with a single non-Hermitian impurity. 

\end{abstract}
\maketitle

\section{Introduction}
\label{sec:intro}
Non-Hermitian (NH) physics has been extensively explored in recent years, uncovering a wide range of phenomena richer than the Hermitian picture and with applications in both classical and quantum regimes \cite{NHbook, NHreview}. 
Significant emphasis has been placed on the systematic study of the topological phases of non-Hermitian systems \cite{NHbook, NHreview, gong, KaShUeSa2019,2022Ayan}, which revealed a breakdown of the correspondence between bulk topology and the appearance of boundary modes  \cite{lee,Xi2018}.
Subsequent theoretical efforts were successful in formulating a generalized bulk-boundary correspondence \cite{BBC, yaowang,KochBBC,KuDw2019,EdKuBe2019,LeLiGo2019,AlVaTo2018,ZirnRefRos,BorgKruchSla,LeeThomale,HerBardReg,Elisabet2020,VyasRoy,NehraRoy} consistent with experiments \cite{HeHoImAbKiMoLeSzGrTh2019,HoHeScSaBrGrKiWoVoKaLeBiThNe2019,GhBrWeCo2019,XiDeWaZhWaYiXu2019}.

A unique feature of NH systems, which has no counterpart in the Hermitian domain, is the accumulation of an extensive number of eigenstates at the boundaries, a phenomenon coined as the \emph{non-Hermitian skin effect} (NHSE)~\cite{yaowang}. 
The phenomenology and origin of this remarkable effect -- including the connection to the existence of non-trivial topological invariants -- remains an active field of current research \cite{Okuma2020, Okuma2022, LinTaiLiLee,Longhi2021,QinMaShen,WangWang,JeonLee,Franca,LiangXie,Yoshida_MSkin2020,Fan}. 
Crystal defects in NH systems \cite{PoBeKuMoSc2015} have also turned out to be an alternative platform to induce the NHSE, even when it is not present under open boundary conditions (OBC)~\cite{NHdisc,NHdisloc, NHdisloc2, NHdisloc3}. 
This extensive theoretical research has effusively led to the exploration of different experimental platforms on which NHSE could be observed \cite{GuGao,LiLeeMuGong}. 
Underpinning the NHSE lies an extreme sensitivity of the spectrum to boundary conditions \cite{BBC,elisabet,Trefethen}, which opens up new potential avenues for sensor applications \cite{NHsensor,NHsensor2,NHsensor3,NHsensor4,NHsensor5}.

Much more well-established is the phenomenon known as Anderson localization \cite{Anderson, Thouless}, present in disorder media. 
In recent years, the interplay between non-Hermiticity and transitions to Anderson localization or non-periodic potentials has also been of increasing interest
\cite{TomKhay,SunLiu2023,WangWang2023,LuOhtShin,ZengChen,SpringKonye,Hui,Hebert, Jiang:2019,LiuZhouChen2021, Longhi2019,Longui20192, OrImu, ChenChengLin,Longui2021, LinLiXiaoWangYiXue, LinLiXiaoWangYiXuePRL}.

On the other hand, the impact of single local impurities on physical properties have also been the subject of much current interest in many other areas of physics~\cite{Stocker:2022,Capizzi:2023,Brauneis:2023,Andreanov:2023,ShenChenQinZhongLee,QinShenLee2023}.

In Ref. \onlinecite{scalefree21}, it was first observed how non-reciprocal impurities in a non-Hermitian Hatano-Nelson chain induced scale-free localized (SFL) states. 
It was also discussed how the variation of the impurity strength could produce transitions between NHSE and SFL, including the counter-intuitive behavior of localization in the direction opposite to the predominant hopping term. 
Moreover, solutions for an on-site impurity in NH Hatano-Nelson and SSH chains were discussed in \cite{impurityproblem2}, showing how the impurity effectively can act as an open boundary condition for the system. 
Similarly, tight connections between the NHSE and the presence of an impurity in one-dimensional lattices were also studied in \cite{impurityproblem,BorgKruchSla}, where an on-site infinite impurity (site vacancy) was considered to develop a method to calculate the eigenstates of the system based on the Green's function method. Further work includes the study of the effect of NH impurities on the properties of Dirac systems \cite{balatsky} or the introduction of topological defects in NH electrical circuits \cite{StegImHelHofLee}.

\begin{figure}[t]
\centering
\includegraphics[width=\columnwidth]{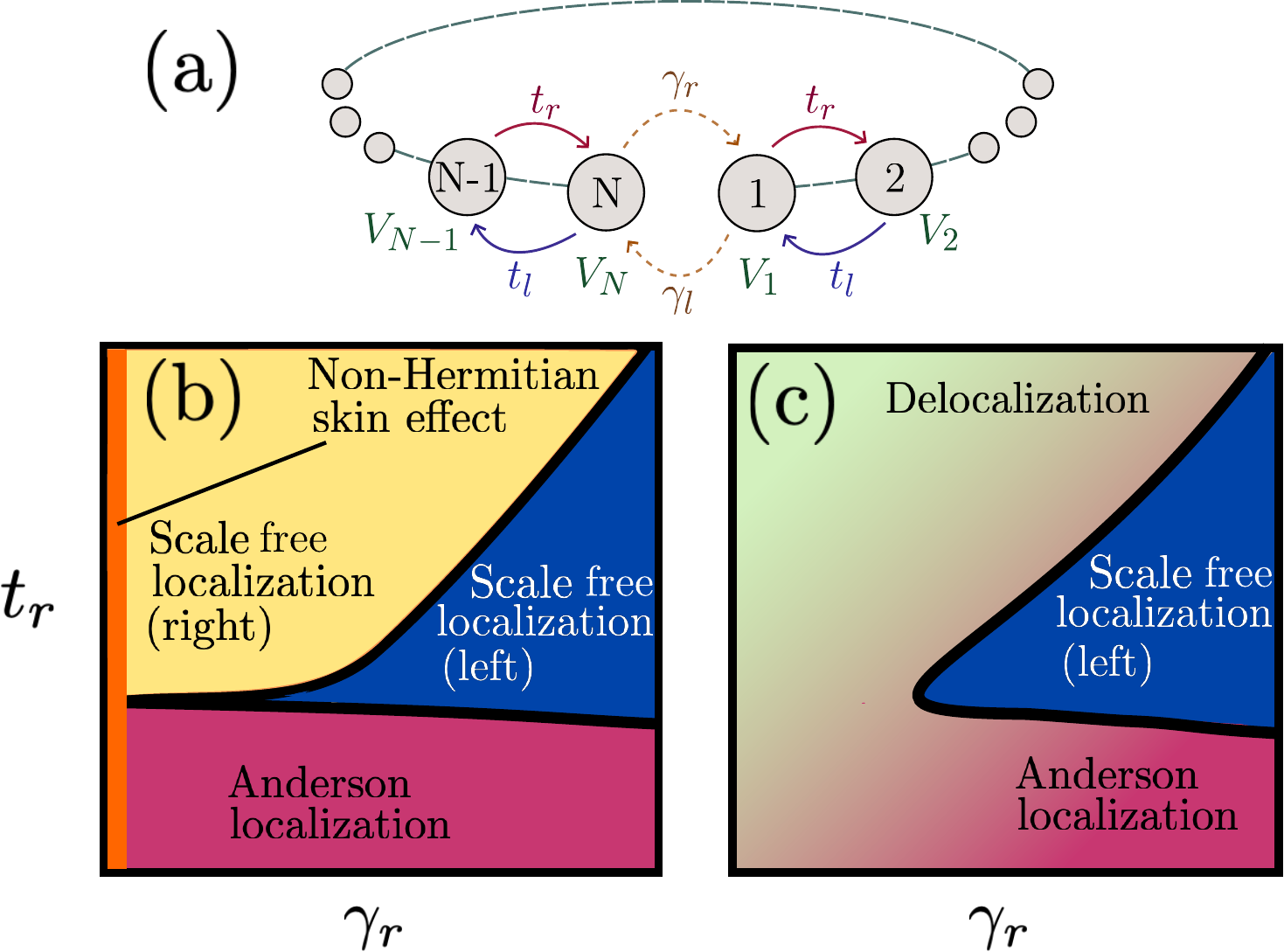}
\caption{(a) Sketch of the system considered in this work, consisting of a chain of sites with hopping $t_r$ (to the right) and $t_l$ (to the left), on-site potentials $V_n$, and impurity hopping between first and last site $\gamma_r$ ($\ket{N} \to \ket{1}$) and $\gamma_l$ ($\ket{1} \to \ket{N}$).
(b)-(c) Sketch of the phase diagram of the model as a function of right hopping $t_r$ and impurity strength $\gamma_r$. In (b) $t_l=0$ and in (c) $t_l=t_r$ (Hermitian hopping).}
\label{fig:summary}
\end{figure}

Here we combine the aforementioned notions of localization by considering a single NH impurity in a one-dimensional tight-binding model subject to on-site disorder (Fig \ref{fig:summary}(a)). 
A rich interplay between different phases of the system is observed, including the NHSE, Anderson localization and the appearance of { scale-free} skin localization  (Fig \ref{fig:summary}(b) and (c)). 
We refer to these as {\it anomalous skin effects} for several reasons. 
First, the { scale-free} localization in the presence of impurities is qualitatively distinct from the NHSE occurring at open boundaries both in that its localization is not dictated by the bulk (Fig \ref{fig:summary}(b)) and that the localization length is not fixed but instead proportional to the system size \cite{scalefree21}. 
Second, in presence of disorder, there is a non-monotonic localization behavior as a function of the hopping terms (Cf. Fig \ref{fig:summary}(b)). 
Third, the key features including a { scale-free} skin localization and non-monotonic dependence of localization as a function of hopping strength persist even in the limit of a NH impurity in an otherwise Hermitian bulk (Fig \ref{fig:summary}(c)). 
To elucidate these points we begin by fully analytically solving a special limit of this system that nevertheless contains key ingredients that, together with the standard phenomenology of Anderson localization, explains our numerically-obtained phase diagrams (Fig \ref{fig:summary}(b) and (c)).

\section{Model}
\label{sec:model}
Throughout this work we consider an extension of the Hatano-Nelson model \cite{HatanoNelson,HatanoNelson2,GuoLiuZhaoLiuChe2021}, a paradigmatic example of a one-dimensional chain in which the non-reciprocity of nearest neighbour hoppings leads to the NHSE under OBC. 
The Hatano-Nelson Hamiltonian reads
\begin{align}
H &=\sum_{n=1}^{N-1}\big( t_r \ket{n+1}\bra{n}+t_l \ket{n}\bra{n+1} \big) \nonumber\\
& \: +\sum_{n=1}^{N}V_n\ket{n}\bra{n}+\gamma_r \ket{1}\bra{N}+\gamma_l \ket{N}\bra{1}, \label{hgeneral}
\end{align}
where $t_r$ and $t_l$ indicate nearest-neighbor hopping respectively to the right and to the left, $V_n$ is an on-site disorder potential that we take from a random uniform distribution $[-V, V]$~\cite{HatanoNelson3}, and $\gamma_r$ and $\gamma_l$ are hopping strengths between the first and last site, which parametrize an impurity. This model is sketched in Fig.~\ref{fig:summary}(a).
In general Eq. (\ref{hgeneral}) is non-Hermitian. 
However, we consider $t_r,t_l,\gamma_r,\gamma_l,V$ to be real, thus $H$ has a real matrix representation, $H=H^*$.
Eigenvalues are consequently real or appear in complex conjugate pairs. 

\section{Results}
\subsection{Solvable limits}
\label{subsec:solvable}
Key insights can already be understood through the study of simple particular cases. 
Solving the Hamiltonian Eq. \eqref{hgeneral} in the simplest non-trivial case when $t_r=t_l=V=\gamma_l=0$ but $\gamma_r\neq 0$, we find an order two exceptional point and $N$-fold energy degeneracy at $E=0$. 
The $N-1$ distinct right eigenstates can be taken as localized at sites $n=1,\ldots,N-1$. 
This solution, however, is highly unstable towards hopping.

As we turn on the hopping, $t_r$, while keeping $t_l=V=\gamma_l=0$, we get a more non-trivial case in which the Hamiltonian reads as 
\begin{eqnarray}
H=\sum_{n=1}^{N-1} t_r\ket{n+1}\bra{n} +\gamma_r \ket{1}\bra{N} .
\label{toymodel}
\end{eqnarray}
Again the eigenspectrum $E_n$ and the (right) eigenvectors  $\ket{\Psi_{R,n}}$ can be found exactly as \cite{GuoLiuZhaoLiuChe2021}
\begin{align}
    E_n &=e^{\frac{2\pi i n}{N}}\, \gamma_r^{\frac{1}{N}} \, t_r^{\frac{1}{N}(N-1)} , \label{toy1} \\ 
    \ket{\Psi_{R,n}} &= \mathcal{N} \left( \sum_{\ell=1}^{N} e^{-\frac{2\pi i n}{N} \ell} \left(\frac{t_r}{\gamma_r}\right)^{\frac{\ell}{N}} \ket{\ell} \right) ,\label{toy2}
\end{align}
with some normalization constant $\mathcal{N}$ and where $n=1,\ldots,N$ labels the eigenstates and their corresponding eigenvalues, which occur on a circle in the complex plane with radius $\gamma_r^{\frac{1}{N}} \, t_r^{\frac{1}{N}(N-1)}$. 
Here it is important to note that the $t_r\rightarrow 0$ limit is not smooth as can be seen by comparing with the first solvable example described at the beginning of the previous paragraph. 

In Eq. (\ref{toy2}) one can readily see that in the limit $\gamma_r \to 0$ all right eigenvectors are completely localized on the right boundary: $\ket{\Psi_{R,n}} \sim \left( \begin{array}{c c c c} 0 & 0 & \cdots & 1 \end{array} \right)^T$.
Dual to this limit, when $\gamma_r \to \infty$, i.e. the boundary coupling is infinitely strong, there also exists perfect boundary localization. 
In the dual limit, however, all right eigenvectors are completely localized at the left boundary:
$\ket{\Psi_{R,n}} \sim \left( \begin{array}{c c c c} 1 & 0 & \cdots & 0 \end{array} \right)^T$ \cite{note}. Note that this is more general: interchanging $t_r$ and $\gamma_r$ changes the localization to the opposite side.

In this toy model we can thus control the localization of the eigenstates through the strength of the impurity $\gamma_r$. Since the expression Eq. (\ref{toy2}) is only valid for a finite $t_{r}$ (analogous results and conclusions would hold for finite $t_{l}$, $\gamma_l$), this clearly implies that adding finite $t_{l/r}$ immediately localizes all eigenstates, even if the added hopping is directed {\it opposite} to the direction of localization. 
The reason for this is essentially that the sites need to be connected in order to make the localization possible. 
The nature of the connections are however not crucial, hence allowing for the counter intuitive phenomenon of localization in the opposite direction compared to the added terms. 
This counter-intuitive effect highlights a different behavior compared to the NHSE, where the localization is towards the leading unbalanced hopping.

We remark that while biorthogonality and simultaneously considering both left and right eigenstates is fundamental for understanding some aspects such the appearance of boundary modes \cite{BBC,NHsensor5}, the phenomenology of the skin effect is readily highlighted by considering the localization properties of (either left or) right states as we do in this work. 

{ Analytical investigation of the toy model in Eq.~\eqref{toymodel} and its experimental realization based on a non-Hermitian circuit platform has been performed in Ref.~\onlinecite{SuGuWaLiRuDuCheZhe}.}

\subsection{Phenomenology beyond the solvable limits}
\label{subsec:beyond-solvable}
From the above exact solutions it is possible to understand the full behavior of the generic model in Eq. \eqref{hgeneral} in detail. To begin with, we keep the maximum nonreciprocity in both the hopping and impurity terms, setting $\gamma_l=t_l=0$, while introducing disorder in the form of a random uncorrelated on-site potential of maximal strength $\pm V$. 
We then scan the $t_r$-$\gamma_r$ parameter space and report the behavior of the eigenstates, as shown in Fig.~\ref{fig:summary}(b).

\begin{figure}[h!]
\centering
\includegraphics[width=\columnwidth]{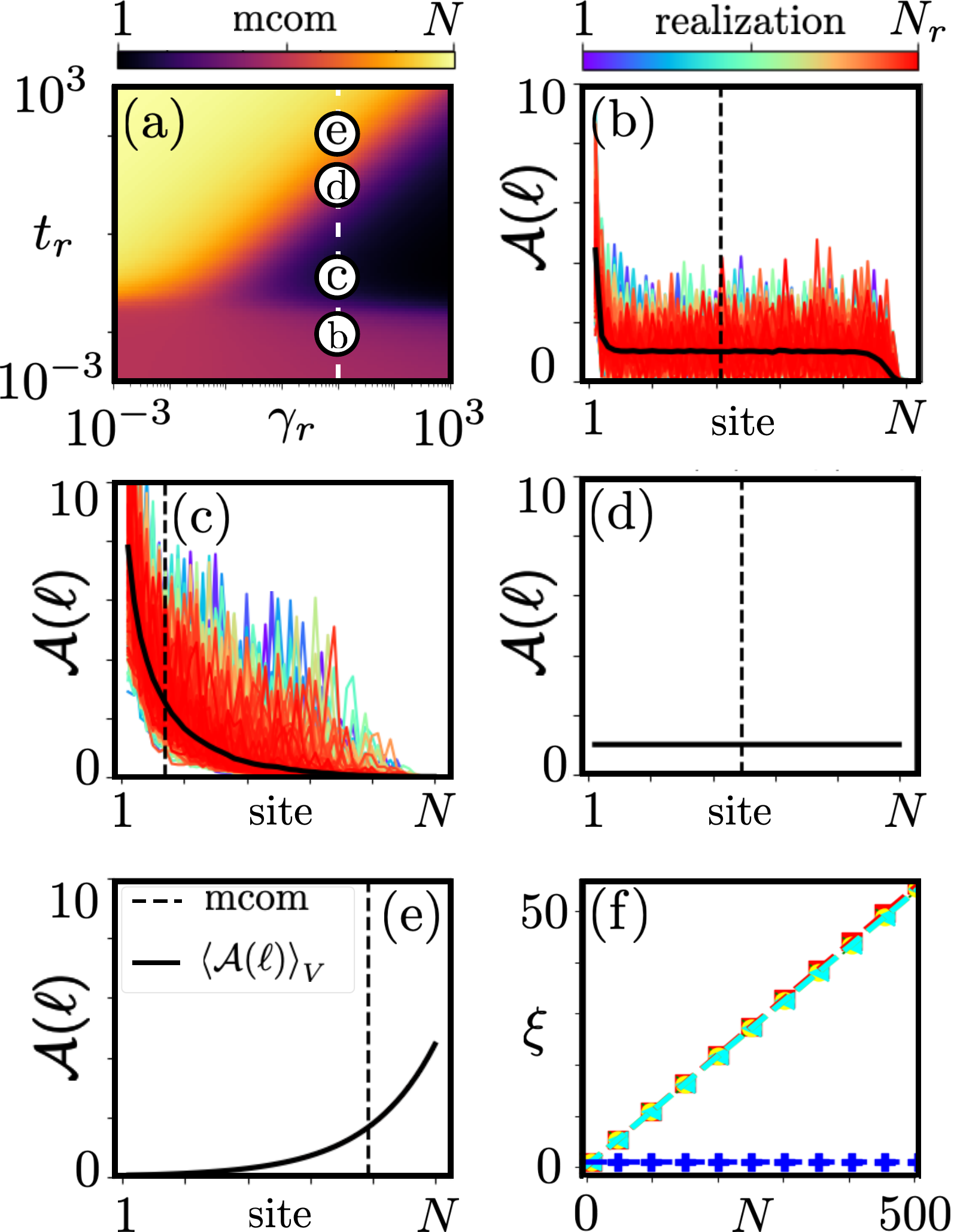}
\caption{(a) mcom as a function of right impurity hopping $\gamma_r$ and right bulk hopping $t_r$ for $V=0.1$, showing the three different phases discussed in the main text.
(b)-(d) mean eigenvector amplitude squared for $N_r=1000$ disorder realizations (colored lines) and disorder-averaged (thick black line) for the four points indicated in panel (a).
The vertical dashed lines indicate the mcom in each case.
(f) Localization length $\xi_{L/R}$ extracted from the { scale-free} left- and right-localized phases (red squares and yellow dots, respectively), compared to the same results obtained from the toy model (green crosses and cyan triangles).
We can observe perfect agreement between the scaling of the toy model and that of the disordered model.
The scaling of the standard NHSE phase at $\gamma_r=0.0$ is also depicted as blue pluses and clearly shows no $N$-dependence. 
The colored dashed lines indicate linear fits to the data points.
}
\label{fig:localization-NH}
\end{figure}

For small values of the impurity strength $\gamma_r$, making the hopping $t_r$ larger (i.e. the non-Hermiticity of the bulk) will gradually increase the localization of the eigenstates to the hopping direction, as expected in the NHSE regime. 
This behavior can be easily quantified by calculating the average eigenstate localization in the form of the mean center of mass (mcom) of the amplitude squared of all eigenvectors $\ket{\Psi_{R,n}}$, averaged over many disorder realizations $N_r$, i.e.
\begin{align}
\left< \mathcal{A}(\ell) \right>_V &= \left< \frac{1}{N} \sum_{n=1}^N |\left< \ell \middle| \Psi_{R,n} \right>|^2 \right>_V , \\
\mathrm{mcom} &= \frac{\sum_{\ell=1}^N \ell \left< \mathcal{A}(\ell) \right>_V}{\sum_{\ell=1}^N \left< \mathcal{A}(\ell) \right>_V} , \label{eq:mcom}
\end{align}
where $\left< \cdot \right>_V$ indicates disorder averaging.
We plot the mcom in Fig.~\ref{fig:localization-NH}(a) over six orders of magnitude for both $t_r$ and $\gamma_r$ and disorder strength $V=0.1$.
From this plot, we can clearly see that in the limit of small $\gamma_r$ and large $t_r$, all eigenstates pile up on the right end of the chain.

We remark that there are other quantities that could be used to probe the localization, but the information they provide should be essentially equivalent to that of the mcom because of the noninteracting nature of the system.
In appendix \ref{app:ipr}, for instance, we show results for the disordered-averaged inverse participation ratio (IPR).
However, while the IPR performs very well in detecting Anderson localization, its measure of the NHSE localization is inferior to that obtained via the mcom.
In appendix \ref{app:biortho}, we show plots of biorthogonal quantities, calculated using both left and right eigenvectors.
Since the left and right eigenstates are localized on opposite boundaries in the NHSE phases, biorthogonal quantities are not useful in determining localization properties and can at most only discriminate whether a phase is Anderson-localized or not.
We also remark that the information provided by the eigenspectrum, such as the complex eigenvalue fraction, is also redundant for our purposes of determining NHSE localization.

More interestingly, increasing the impurity strength causes the system to exhibit a different behavior.
For small values of the right hopping $t_r$, the disorder introduced through the on-site potential will always dominate, leading the eigenstates to localize on the basis of Anderson localization.
This corresponds to a $\mathrm{mcom} \approx N/2$ in Fig.~\ref{fig:localization-NH}(a) (purple region).
However, at sufficiently large values of $t_r$, we can distinguish two different behaviors of the system when $\gamma_r$ is ramped up to progressively larger values.

At first, when $t_r \ll \gamma_r$, the increase of the right hopping $t_r$ causes the eigenstates to pile up to the \textit{left} of the chain, following the counterintuitive picture already observed in the toy model at $V=0$.
In Fig.~\ref{fig:localization-NH}(a), this is signalled by $\mathrm{mcom} \approx 1$ (dark-colored region).
The localization is exponential, i.e. the disorder-averaged mean amplitude squared of all eigenvectors can be very well fitted by $\left< \mathcal{A}(\ell) \right>_V =  A_L \exp \left( -\frac{\ell}{\xi_L} \right)$ with the localization length $\xi_L$ and some amplitude $A_L$.

Upon increasing the right hopping beyond $t_r \gtrapprox \gamma_r$, however, the localization is rapidly inverted and all eigenstates pile up again towards the \emph{right} end of the chain (next to the impurity) as observed for small $\gamma_r$ and large $t_r$.
Again, the localization is exponential, with the form $\left< \mathcal{A}(\ell) \right>_V =  A_R \exp \left( -\frac{N-\ell}{\xi_R} \right)$.
The progression from Anderson localized, to left-localized, to right localized skin effects is depicted explicitly in panels (b)-(e) of Fig.~\ref{fig:localization-NH}, where the mean amplitude squared of all eigenvectors is plotted for $N_r=1000$ disorder realizations.

In both skin localized phases, the localization length $\xi_{L/R}$ is proportional to the system size $N$, as can be seen in Fig.~\ref{fig:localization-NH}(f).
We find that $\xi_L \simeq \xi_R$ over many orders of magnitude in $N$, indicating that the underlying scaling is the same for both left- and right-localization.
Furthermore, the scaling matches perfectly with what we can obtain exactly for the clean limit of the toy model \eqref{toymodel}, i.e. when $V=0.0$, $t_l=r_l=0.0$.
This agreement demonstrates that the core mechanism for the localization physics lies within the interplay between hopping and impurity, but is stabilized by the disorder to a proper phase. 
This linear $N$-dependence is to be starkly contrasted with the standard NHSE occurring at $\gamma_r=0.0$, where the localization length remains instead constant for any value of $N$ (green markers and line).
This unique phenomena of \textit{{ scale-free}} eigenstates, representing an \textit{anomalous skin effect} are always accompanied by the emergence of complex eigenspectrum, as observed in \cite{scalefree21}. 

We remark that the nonmonotonicity in the localization arises not from the NHSE phase, but rather from the Anderson localized phase. 
The disorder element is therefore crucial: at low values of disorder the nonmonotonicity namely disappears completely and the same behaviour is observed in presence of a periodic potential, see the supplementary material for details~\cite{supmat}. 
A heuristic explanation for the lack of nonmonotonicity in the periodic potential case is that the eigenstates remain delocalized in absence of the NH impurity, and hence immediately feel its presence when turned on.   
We also note that larger values of $V$ will lead to qualitatively similar phase diagrams to the one shown in Fig.~\ref{fig:localization-NH}, but where the appearance of the left-localized phase is shifted to larger values of the parameters.
Similarly, nonzero values of $t_l$ and $\gamma_l$ might shift the detailed shape of the phase diagram, but do not alter the generic features we presented.

\subsection{Analytical mcom in the clean case}
\label{subsec:anal-mcom}
\begin{figure}[h!]
\centering
\includegraphics[width=0.9\columnwidth]{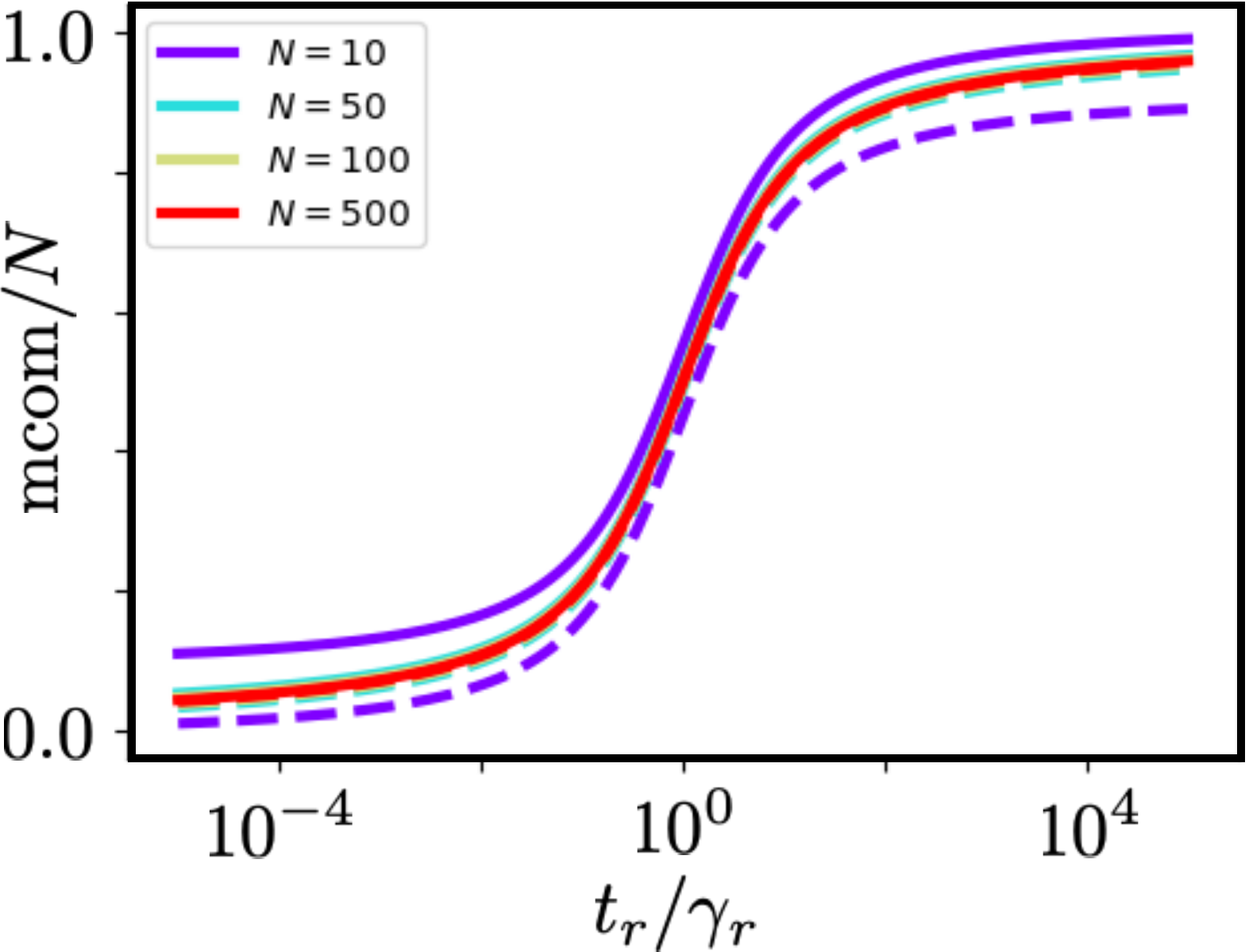}
\caption{{
 Behavior of the mean center-of-mass as a function of the ratio $t_r/\gamma_r$ and for increasing values of chain length $N$ in the clean case $V=0$. Solid lines are calculated analytically with formula Eq. \eqref{eq:mcom-anal}, while dashed lines correspond to numerical results.
}
}
\label{fig:mcom-anal}
\end{figure}

By using the exact solutions of the toy model, Eq.~\eqref{toy2}, it is possible to obtain an analytical expression for the mcom in the limit of zero disorder $V=0$ (see supplementary material for detailed derivation):
\begin{equation}
    \mathrm{mcom} \big|_{V=0}  = N + \frac{1}{1- \left(\frac{t_r}{\gamma_r} \right)^{2/N}} + \frac{N}{\left(\frac{t_r}{\gamma_r} \right)^2 - 1}.
    \label{eq:mcom-anal}
\end{equation}
Fig.~\ref{fig:mcom-anal} shows the analytical mcom normalized with the chain length for various values of $N$.
The analytical result showcases the change in localization from the left to the right of the chain as a function of the ratio $t_r/\gamma_r$, and is corroborated by the numerical results.
Furthermore, we can appreciate how in the limit of large $N$, all curves fall on top of each other, indicating scale-free localization.

\subsection{Hermitian hopping case}
\label{subsec:hermitian}
As we have explored in the solvable model, large values of the right impurity strength $\gamma_r$ dominate the system behavior, localizing eigenstates to the \textit{left}. 
This means that the increase of the hopping in the direction of the localization (i.e. $t_l$) will have no effect on the localization behavior, i.e. the { scale-free} localization to the \textit{left} will prevail. 
Thus adding such hopping to the solvable model Eq. \eqref{toymodel} to make it Hermitian in the bulk, i.e. $t_l=t_r$ preserves the localization property of Eqs. \eqref{toy1}-\eqref{toy2}.
This sheds light on the provenance of the { scale-free} localization induced by local non-Hermitian impurities in otherwise Hermitian systems as noticed very recently in Refs.~\onlinecite{scalefree23,scalefree232}.
In the Hermitian case, however, neither { scale-free} localization to the right nor NHSE is observed, as would be expected. 

\begin{figure}[h!]
\centering
\includegraphics[width=\columnwidth]{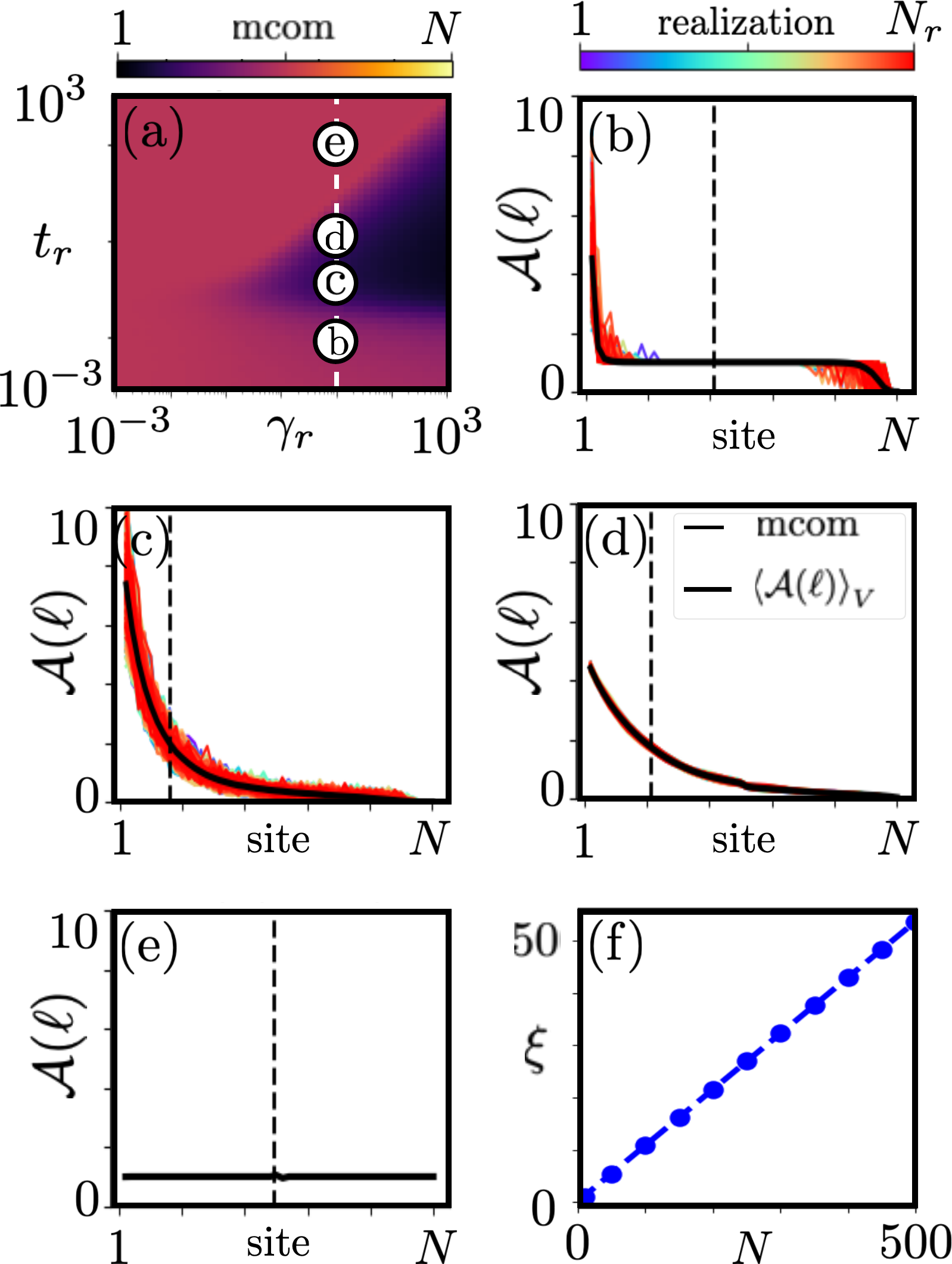}
\caption{(a) mcom as a function of right impurity hopping $\gamma_r$ and Hermitian bulk hopping $t_l=t_r$ for $V=0.1$.
(b)-(d) mean eigenvector amplitude squared for $N_r=1000$ disorder realizations (colored lines) and disorder-averaged (thick black line) for the four points indicated in panel (a).
The vertical dashed lines indicate the mcom in each case.
(f) Localization length $\xi_{L}$ extracted from the { scale-free} left-localized phase.
The dashed line indicate a linear fit to the data points.
}
\label{fig:localization-herm}
\end{figure}

The sketch of the phase diagram in the Hermitian case is displayed in Fig. \ref{fig:summary}(c).
A more detailed description of the localization properties is instead presented in Fig.~\ref{fig:localization-herm}. 
In panel (a), the phase diagram is obtained by means of the mcom, where we can clearly see the absence of localization at right boundary.
Note that from the perspective of the mcom, the Anderson localized phase and the phase where the hopping dominates (for $t_r>\gamma_r > V$) cannot be distinguished from each other. 
In this case, the inverse participation ratio could be used as additional tool (see appendix \ref{app:ipr}).
However, here we are mainly interested in scale-free localization behavior.

Panels (b)-(e) show the mean amplitude squared of all eigenvectors $N_r=1000$ disorder realizations, and increasing value of $t_r$ corresponding to the points displayed in panel (a). 
While the left localization is achieved upon increasing the right hopping to values comparable to but smaller than $\gamma_r$, further increasing $t_r$ simply delocalizes the eigenstates across the entire chain.
Finally, panel (f) shows the localization length $\xi_L$ obtained by fitting the disorder-averaged mean amplitude squared with an exponential function. 
Again, we observe a linear behavior with the system size $N$ indicating { scale-free} localization, and indeed with the same slope observed for the non-Hermitian hopping case.

\section{Conclusions} 
\label{sec:conclusions}
We have explored anomalous skin localization phenomena induced by non-Hermitian impurities. 
On a basic level this showcases fundamental properties of NH spectra \cite{Trefethen} and their associated novel eigenvector properties \cite{ChalkerMehlig}. 
At the same time it relates directly to experimental realities in a variety of systems ranging from robotic metamaterials \cite{GhBrWeCo2019} and electrical circuits \cite{HeHoImAbKiMoLeSzGrTh2019,HoHeScSaBrGrKiWoVoKaLeBiThNe2019} to optical systems \cite{XiDeWaZhWaYiXu2019,funnel,photonics} in which the standard NHSE has already been observed.   

While the effect of impurities in NH one-dimensional tight-binding models has been explored in previous works, here we have added several important aspects.
First, we have identified a minimal analytically solvable model that exhibits anomalous skin effects. 
The solution of our model do not involve any approximations and yields all eigenenergies and eigenstates at any finite size. 
This model highlights a previously overlooked non-analytic weak hopping limit and a counterintuitive non-monotonic relation between (directional) hopping and localization. 
Second, we have established that, by adding bulk disorder, the aforementioned non-monotonic behavior is promoted from a highly fine-tuned point in parameter space into a generic and stable phenomenon. 
Third, we have shown that adding bulk hopping in the direction of the scale-free localization cannot undo it (in fact it leaves it unchanged) hence explaining why a single non-Hermitian impurity can also induce a scale-free skin localization in an otherwise Hermitian system.    

These results corroborate the potential for harnessing impurities for local sensing and control of a large class of effectively non-Hermitian systems. 
That the phenomenology also extends to systems that are Hermitian in the bulk, which follows transparently from perturbing away our solvable model, further extends the scope of these insights.  

Finally, very shortly before posting this work, several pre-prints pointed out the possibility of inducing scale-free localization through NH impurities in a Hermitian bulk based on tight-binding models distinct from ours \cite{scalefree23,scalefree232}, as well as by solving dissipating spin chains in the thermodynamic limit using Bethe ansatz \cite{bethescalefree}.

{\it Acknowledgements.--}
The authors are supported by the Swedish Research Council (VR consolidator grant 2018-00313), the Knut and Alice Wallenberg Foundation (KAW) via the Wallenberg Academy Fellows program (2018.0460) and the G\"{o}ran Gustafsson Foundation for Research in Natural Sciences and Medicine. This work was partially supported by the Knut and Alice Wallenberg Foundation through the Wallenberg Centre for Quantum Technology (WACQT).

\appendix

\begin{figure*}[!t]
\centering
\includegraphics[width=0.7\textwidth]{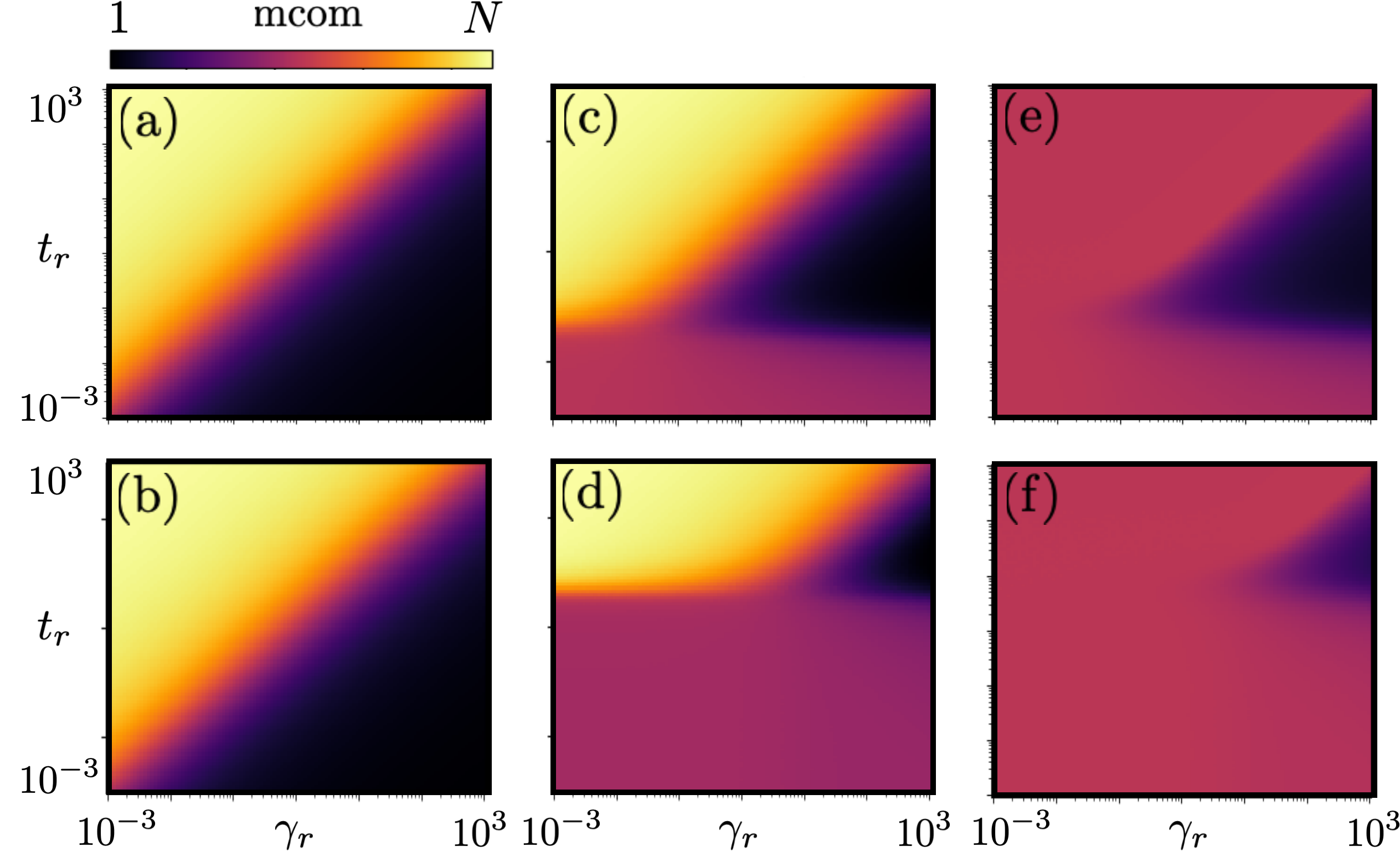}
\caption{Mean eigenvector center of mass for a generalized Hatano-Nelson chain of $N=50$ sites under different conditions, plotted as a function of $\gamma_r$ and $t_r$. 
(a) Clean limit (toy model) with $V=0$, $t_l=\gamma_l=0$, obtained analytically from \eqref{eq:mcom-anal} (numerical results are identical).
(b) Alternating on-site potential of strength $V=10.0$, $t_l=\gamma_l=0$.
(c) Disordered on-site potential of magnitude $V=0.1$, $t_l=\gamma_l=0$.
(d) Disordered on-site potential of magnitude $V=10.0$, $t_l=\gamma_l=0$.
(e) Disordered on-site potential of magnitude $V=0.1$, $t_l=t_r$ (Hermitian hopping), $\gamma_l=0$.
(f) Disordered on-site potential of magnitude $V=10.0$, $t_l=t_r$ (Hermitian hopping), $\gamma_l=0$.
}
\label{fig:phase-diagrams-supmat}
\end{figure*}

\begin{figure*}[!t]
\centering
\includegraphics[width=0.7\textwidth]{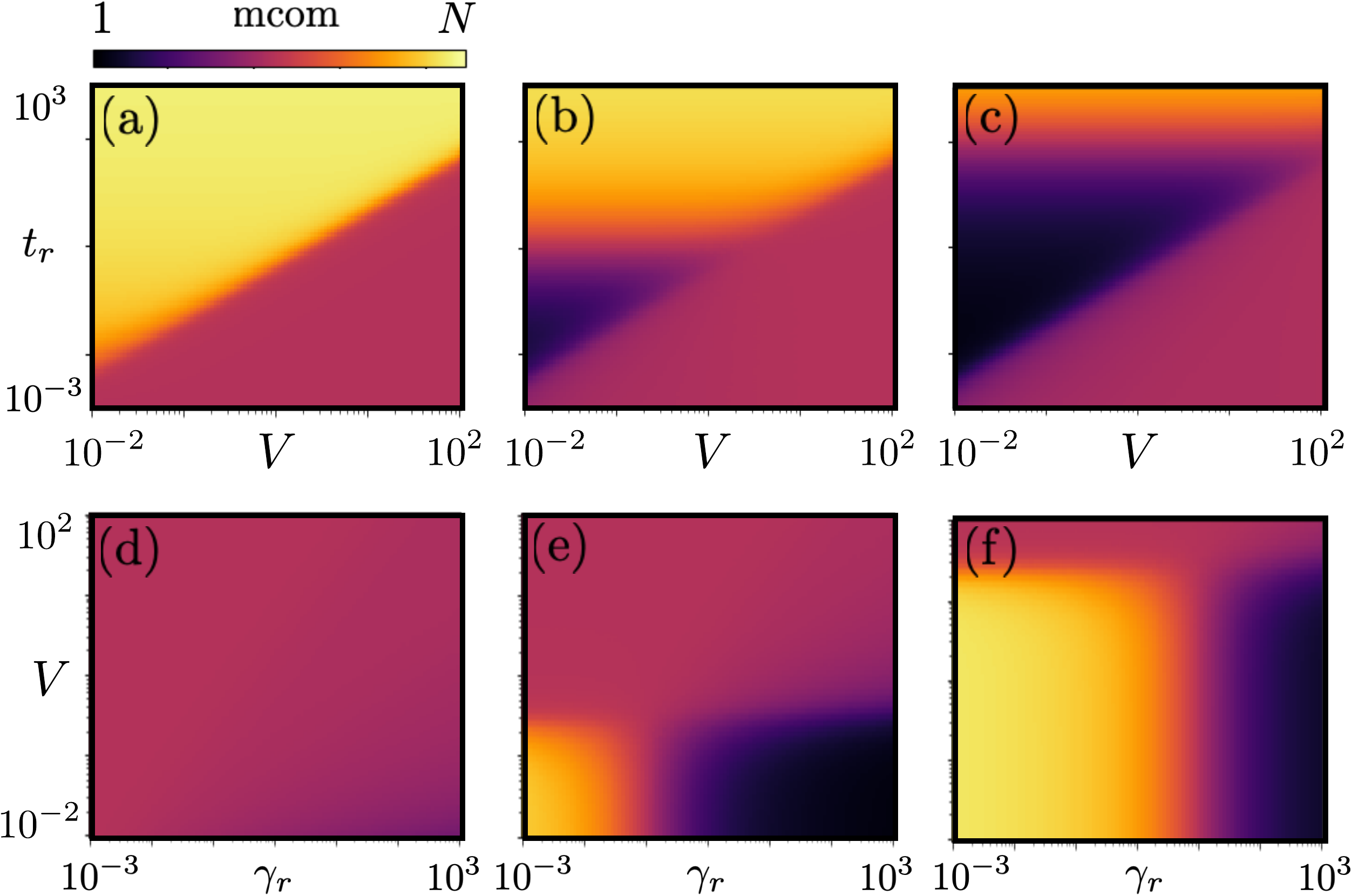}
\caption{
Mean eigenvector center of mass for a generalized Hatano-Nelson chain of $N=50$ sites as a function of $t_r$, $\gamma_r$, and $V$.
(a)-(c) Dependence on $V$ and $t_r$ at fixed $\gamma_r$ for (a) $\gamma_r=0.001$, (b) $\gamma_r=1.0$, (c) $\gamma_r=100$.
(d)-(f) Dependence on $V$ and $\gamma_r$ at fixed $t_r$ for (d) $t_r=0.001$, (e) $t_r=0.1$, (f) $t_r=10$.}
\label{fig:phase-diagrams-supmat2}
\end{figure*}

\section{Localization as a function of on-site potential}
\label{app:V-dependence}
In this section, we summarize the behavior of the mean center of mass (mcom) Eq.~\eqref{eq:mcom} as a function of different on-site potential strengths.
This quantity was used in the main text as an effective order parameter to classify the system into phases with different localization properties.
We have calculated the mcom as a function of the parameters $t_r$ and $\gamma_r$ for $N=50$ sites and $t_l=\gamma_l=0.0$. 
The results are shown in Fig.~\ref{fig:phase-diagrams-supmat}.
Panel (a) shows the results for the clean limit given by the toy model of Eq.~\eqref{toymodel}.
As discussed in the main text, the nonmonotonic behavior in the eigenstate (left) localization is not present in this limit: for any value of $\gamma_r > t_r$, increasing $t_r$ will only localize the eigenstates more to the right.

The same behavior is observed also when we add an alternating on-site potential to the toy model, i.e. for
$V_{2n-1} = V$, $V_{2n} = -V$, $n=1, \dots, N$ in Eq.~\eqref{hgeneral}.
Remarkably, the localization properties are identical to those of the toy model, as shown in Fig.~\ref{fig:phase-diagrams-supmat}(b). 

In order to introduce a nonmonotonic behavior in the eigenstate localization, we need to add a disordered potential as explained in the main text.
This leads to the emergence of an Anderson localized phase at small values of the right hopping $t_r$.
For large enough values of the impurity hopping $\gamma_r$, it is then possible to \emph{left}-localize the eigenstates by increasing the hopping to the \emph{right}.
We find that this qualitative picture persists for all values of the disorder potential strength $V$, whereas $V$ controls the onset of the left-localization as seen in Fig.~\ref{fig:phase-diagrams-supmat}(c) and (d).

In the Hermitian hopping regime $t_l=t_r$, it is still possible to control the eigenstate localization by increasing $t_r$ and large values of $\gamma_r$, much like in the non-Hermitian case.
This feature is shown in Fig.~\ref{fig:phase-diagrams-supmat}(e) and (f).
Here, however, the right-localized phase disappears completely and the localization is only possible on the left end of the chain.
Again, increasing the value of $V$ will not change this qualitative picture, but will simply shift the left-localized phase to larger values of $t_r$ and $\gamma_r$.

\begin{figure}
\centering
\includegraphics[width=0.7\columnwidth]{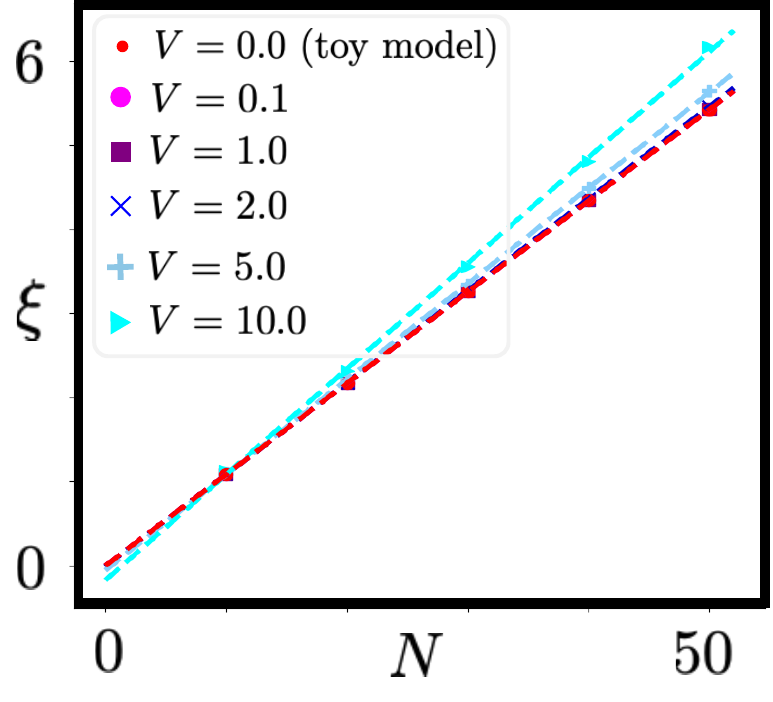}
\caption{Scaling of the (left) localization length as a function of chain length $N$ up to $N=50$ for increasing values of the disorder strength $V$.
The other parameters are $t_l=0.0$, $t_r=10^3$, $\gamma_r=10$, $\gamma_l=0.0$.}
\label{fig:xi-V}
\end{figure}

A comprehensive analysis of the localization behavior as a function of all system parameters ($t_r$, $\gamma_r$, and $V$) is shown in Fig.~\ref{fig:phase-diagrams-supmat2}.
The picture that comes out of these phase diagrams is consistent with the expectation that 
when $V$ is the dominant energy scale, the states are Anderson localized with a mcom of around $N/2$; 
when $t_r$ is the dominant energy scale, the states localize on the right end of the chain; 
when $\gamma_r$ is the dominant energy scale, the states localize on the left end of the chain.

The strength of the disorder can also impact how the localization length scales with system size.
More precisely, we find that when $V$ becomes comparable to the hopping strength $t_r$, the slope of $\xi(N)$ begins to increase and deviate from the clean limit of the toy model already at values $N\sim50$, as shown in Fig.~\ref{fig:xi-V}.
This can be understood as a consequence of the shift of the localized phases towards larger values of $t_r$ and $\gamma_r$ induced by increasing values of $V$, already seen in Fig.~\ref{fig:phase-diagrams-supmat}(c) and (d).
When $\gamma_r$ is kept fixed at a large enough value such that the system lies in the left-localized phase (dark area in Fig.~\ref{fig:phase-diagrams-supmat}), increasing $V$ will progressively shift the mcom towards more central values (dark purple region in Fig.~\ref{fig:phase-diagrams-supmat}), affecting the scaling of the localization length in the process.
This behavior indicates that disorder might be used as an additional lever in controlling the scale-free localization observed in the generalized Hatano-Nelson model.

\section{Inverse participation ratio}
\label{app:ipr}
\begin{figure}[t!]
\centering
\includegraphics[width=\columnwidth]{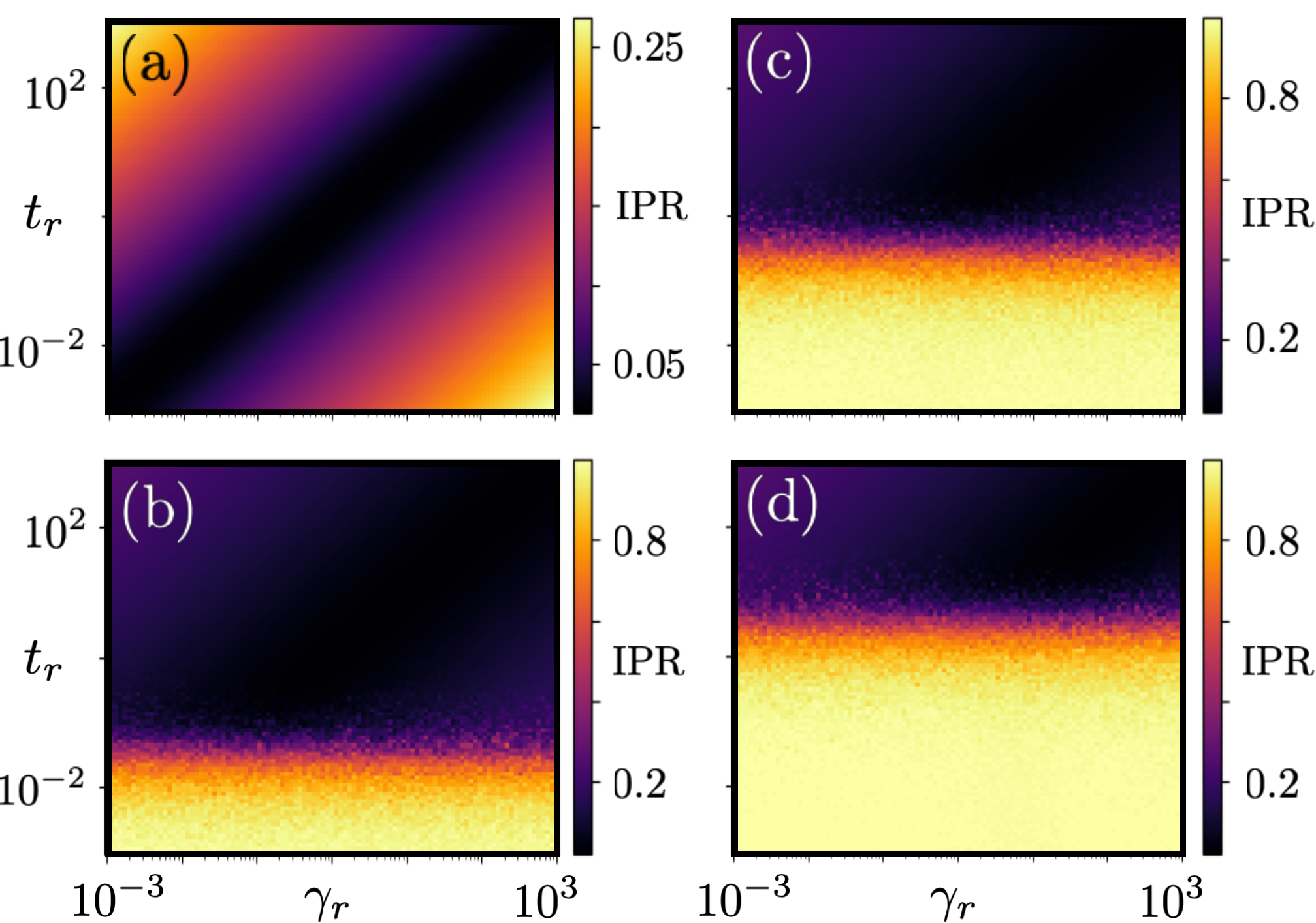}
\caption{
Disordered-average IPR in the $(\gamma_r, t_r)$-parameter space for increasing values of disorder potential $V$.
(a) $V=0$.
(b) $V=0.1$.
(C) $V=1.0$.
(D) $V=10.0$.
}
\label{fig:ipr}
\end{figure}

In this section, we present numerical results for the disordered-averaged inverse participation ratio, or IPR, defined as
\begin{equation}
    \mathrm{IPR} = \left< \frac{1}{N} \sum_{n=1}^N \sum_{m=1}^N |\Psi_{R,n}(m)|^4 \right>_V,
\end{equation}
where we have additionally assumed that the eigenvectors $\ket{\Psi_{R,n}}$ are normalized to one, i.e. $\sum_{m=1}^N |\Psi_{R,n}(m)|^2 = 1$.
The IPR is a standard measure of Anderson localization in noninteracting systems.
Our numerical results are shown in Fig.~\ref{fig:ipr}.
As we can see from the panels, the IPR is a very good diagnostic tool to discriminate the Anderson localized phase from the anomalous NHSE.
However, it performs poorly when distinguishing the NHSE phase from each other, because it only gives a measure of the total localization, and not of the position of the localization.
Even at $V=0$ -- in Fig.~\ref{fig:ipr}(a) -- while the two anomalous NHSE phases can be distinguished, the transition line is completely obscured.
In contrast, the mcom gives a much clearer signal.

\section{Derivation of the analytical expression for the mcom in the clean case}
\label{app:anal-mcom}
Here we derive Eq.~\eqref{eq:mcom-anal} for clean case $V=0$.
We start from the definition of the mcom, Eq.~\eqref{eq:mcom}, which for zero disorder is simply
\begin{equation}
   \mathrm{mcom} \big|_{V=0} = \frac{\frac{1}{N} \sum_{\ell,n=1}^N \ell | \braket{\ell | \Psi_{R,n}} |^2}{\frac{1}{N} \sum_{\ell,n=1}^N | \braket{\ell | \Psi_{R,n}} |^2} .
\end{equation}
By inserting the analytical result for the right eigenvector, Eq.~\eqref{toy2}, we simplify the expression further:
\begin{widetext}
\begin{align}
   \mathrm{mcom} \big|_{V=0} 
   &= \frac{\sum_{\ell,n=1}^N \ell |\mathcal{N}(n) \sum_{\ell'=1}^N e^{-2\pi i n \ell'/N} \left(\frac{t_r}{\gamma_r} \right)^{\ell'/N} \braket{\ell | \ell'} |^2}{\sum_{\ell,n=1}^N  |\mathcal{N}(n)  \sum_{\ell'=1}^N e^{-2\pi i n \ell'/N} \left(\frac{t_r}{\gamma_r} \right)^{\ell'/N} \braket{\ell | \ell'} |^2} 
= \frac{\sum_{\ell=1}^N \ell \theta^{\ell}}{ \sum_{\ell=1}^N  \theta^{\ell}},
\end{align}
\end{widetext}
where we have used $\braket{\ell|\ell'} = \delta_{\ell, \ell'}$ and introduced the ratio $\theta \equiv \left( \frac{t_r}{\gamma_r} \right)^{2/N}$.
The above geometric sums can finally be evaluated exactly with the formulas
\begin{align}
    \sum_{k=0}^N \theta^k &= \frac{1 - \theta^{N+1}}{1 - \theta}, \\
    \sum_{k=0}^N k\theta^k &= \frac{\theta ( N \theta^{N+1} - (N+1) \theta^N + 1)}{(1 - \theta)^2} ,
\end{align}
to yield the compact expression
\begin{align}
   \mathrm{mcom} \big|_{V=0}  = N + \frac{1}{1-\theta} + \frac{N}{\theta^N - 1}.
\end{align}

\section{Biorthogonal quantities}
\label{app:biortho}
\begin{figure}[t!]
\centering
\includegraphics[width=\columnwidth]{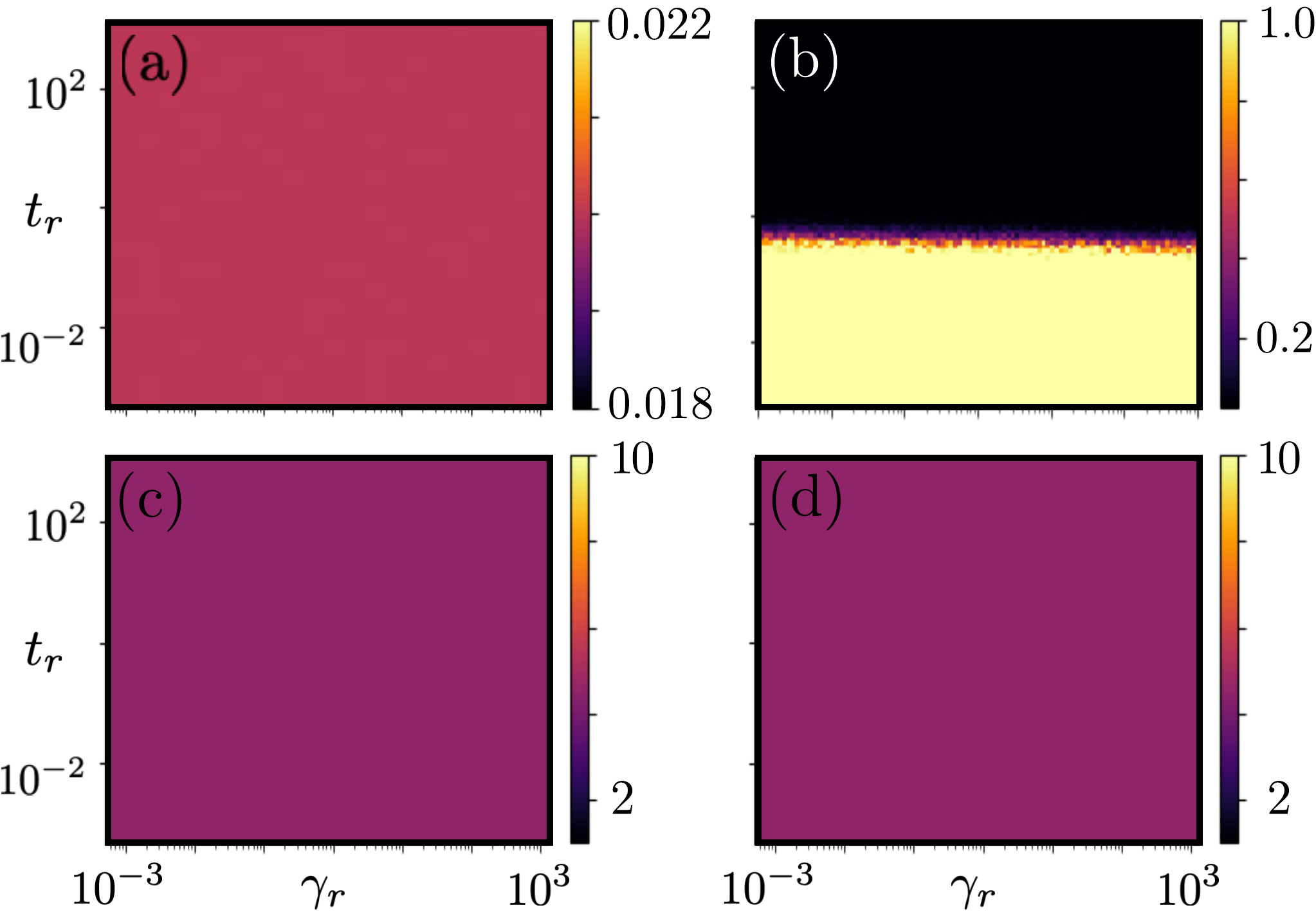}
\caption{
(a)-(b) Biorthogonal inverse participation ratio as a function of $\gamma_r$ and $t_r$ for (a) $V=0.0$ and (b) $V=1.0$.
(c)-(d) Biorthogonal polarization as a function of $\gamma_r$ and $t_r$ for (a) $V=0.001$ and (b) $V=1.0$.
}
\label{fig:biortho}
\end{figure}

Since we are dealing with a non-Hermitian system, one might wonder whether using biorthogonal quantities constructed from both left and right eigenvectors would be more appropriate.
As we can see in Fig.~\ref{fig:biortho}(a)-(b), the biorthogonal generalization of the IPR~\cite{Xiao:2022}
\begin{equation}
    \mathrm{IPR}_{bi} = \left< \frac{1}{N} \sum_n \frac{\sum_{j} |\psi_{L,n}(j)|^2|\psi_{R,n}(j)|^2}{\left( \sum_{j} |\psi_{L,n}(j)||\psi_{R,n}(j)| \right)^2} \right>_V
\end{equation}
gives exactly the same information of the usual IPR and is able to only discriminate between Anderson localization and NHSE.
The disorder-averaged biorthogonal polarization, instead, defined as~\cite{BBC, Elisabet2020}
\begin{equation}
\mathcal{P} = \left< \sum_m \left( 1 - \frac{1}{N} \sum_n n \left<\psi_{L,m} \middle| n \right>\left< n \middle| \psi_{R,m} \right> \right) \right>_V
\end{equation}
with $\ket{\psi_{L/R,m}}$ the $m$-th left (right) eigenstate, is trivial in every phase as shown in Fig.~\ref{fig:biortho}(c)-(d).
This can be explained by the fact that the left and right eigenstates are localized on opposite boundaries, and thus constructing biorthogonal overlaps completely smears out any information about boundary localization.
Since we are mainly interested in the localization properties of the left and right eigenstates separately, examining quantities that only stem from left or only from right eigenstates is the appropriate way to extract information about NHSE phases.


\begin{thebibliography}{100}

\bibitem{NHbook}
Y. Ashida, Z. Gong, and M. Ueda, {\em Non-Hermitian Physics}, \href{https://dx.doi.org/10.1080/00018732.2021.1876991}{Advances in Physics 69, {\bf 249} (2020)}.

\bibitem{NHreview}
 E.J. Bergholtz, J.C. Budich, and F.K. Kunst, {\em Exceptional topology of non-Hermitian systems}, \href{https://journals.aps.org/rmp/abstract/10.1103/RevModPhys.93.015005}{Rev. Mod. Phys. {\bf 93}, 015005 (2021)}.

\bibitem{gong}
Z. Gong, Y. Ashida, K. Kawabata, K. Takasan, S. Higashikawa, and M. Ueda, {\em Topological Phases of Non-Hermitian Systems}, \href{https://journals.aps.org/prx/abstract/10.1103/PhysRevX.8.031079}{Phys. Rev. X {\bf 8}, 031079 (2018)}.

\bibitem{KaShUeSa2019} K. Kawabata, K. Shiozaki, M. Ueda, and M. Sato, \textit{Symmetry and topology in non-Hermitian physics}, \href{https://journals.aps.org/prx/abstract/10.1103/PhysRevX.9.041015}{Phys. Rev. X \textbf{9}, 041015 (2019)}.

\bibitem{2022Ayan}
A. Banerjee, R. Sarkar, S. Dey, and A. Narayan.
\textit{Non-Hermitian Topological Phases: Principles and Prospects}. \href{https://arxiv.org/abs/2212.06478}{arXiv:2212.06478}.


\bibitem{lee}
T.E. Lee, {\em Anomalous Edge State in a Non-Hermitian Lattice}, \href{https://journals.aps.org/prl/abstract/10.1103/PhysRevLett.116.133903}{Phys. Rev. Lett. {\bf 116}, 133903 (2016)}.

\bibitem{Xi2018} Y. Xiong, \textit{Why does bulk boundary correspondence fail in some non-Hermitian topological models}, \href{https://iopscience.iop.org/article/10.1088/2399-6528/aab64a}{J. Phys. Commun. \textbf{2}, 035043 (2018)}.


\bibitem{BBC}
F.K. Kunst, E. Edvardsson, J.C. Budich, and E.J. Bergholtz, {\em Biorthogonal Bulk-Boundary Correspondence in non-Hermitian Systems}, \href{https://journals.aps.org/prl/abstract/10.1103/PhysRevLett.121.026808}{Phys. Rev. Lett. {\bf 121}, 026808 (2018)}.

\bibitem{yaowang}
S. Yao and Z. Wang, {\em Edge states and topological invariants of non-Hermitian systems}, \href{https://doi.org/10.1103/PhysRevLett.121.086803}{Phys. Rev. Lett. {\bf 121}, 086803 (2018)}.

\bibitem{KochBBC} R. Koch  and J.C. Budich, {\em Bulk-boundary correspondence in non-Hermitian systems: Stability analysis for generalized boundary conditions}, \href{https://link.springer.com/article/10.1140/epjd/e2020-100641-y}{Eur. Phys. J. D \textbf{74}, 70 (2020)}.

\bibitem{KuDw2019} F.K. Kunst and V. Dwivedi, \textit{Non-Hermitian systems and topology: A transfer-matrix perspective}, \href{https://journals.aps.org/prb/abstract/10.1103/PhysRevB.99.245116}{Phys. Rev. B \textbf{99}, 245116 (2019)}.

\bibitem{EdKuBe2019} E. Edvardsson, F.K. Kunst, and E.J. Bergholtz, \textit{Non-Hermitian extensions of higher-order topological phases and their biorthogonal bulk-boundary correspondence}, \href{https://journals.aps.org/prb/abstract/10.1103/PhysRevB.99.081302}{Phys. Rev. B \textbf{99}, 081302(R) (2019)}.

\bibitem{Elisabet2020} E. Edvardsson, F.K. Kunst, T. Yoshida, and E.J. Bergholtz, \textit{Phase transitions and generalized biorthogonal polarization in non-Hermitian systems}, \href{https://journals.aps.org/prresearch/abstract/10.1103/PhysRevResearch.2.043046}{Phys. Rev. Research {\bf 2}, 043046 (2020)}.

\bibitem{LeLiGo2019} C.H. Lee, L. Li, and J. Gong, \textit{Hybrid higher-order skin-topological modes in nonreciprocal systems}, \href{https://journals.aps.org/prl/abstract/10.1103/PhysRevLett.123.016805}{Phys. Rev. Lett. \textbf{123}, 016805 (2019)}.

\bibitem{AlVaTo2018} V.M. Martinez Alvarez, J.E. Barrios Vargas, and L.E.F. Foa Torres, \textit{Non-Hermitian robust edge states in one dimension: Anomalous localization and eigenspace condensation at exceptional points}, \href{https://journals.aps.org/prb/abstract/10.1103/PhysRevB.97.121401}{Phys. Rev. B \textbf{97}, 121401(R) (2018)}.

\bibitem{ZirnRefRos} H.G. Zirnstein, G. Refael, and B. Rosenow, \textit{Bulk-Boundary Correspondence for Non-Hermitian Hamiltonians via Green Functions}, \href{https://journals.aps.org/prl/abstract/10.1103/PhysRevLett.126.216407}{Phys. Rev. Lett. \textbf{126}, 216407 (2021)}.

\bibitem{BorgKruchSla} D.S. Borgnia, A.J. Kruchkov, and R.J. Slager, \textit{Non-Hermitian Boundary Modes and Topology}, \href{https://journals.aps.org/prl/abstract/10.1103/PhysRevLett.124.056802}{Phys. Rev. Lett. \textbf{124}, 056802 (2020)}.

\bibitem{LeeThomale} C.H. Lee and R. Thomale, \textit{Anatomy of skin modes and topology in non-Hermitian systems}, \href{https://journals.aps.org/prb/abstract/10.1103/PhysRevB.99.201103}{Phys. Rev. B \textbf{99}, 201103(R) (2019)}.

\bibitem{HerBardReg} L.Herviou, J.H. Bardarson, and N. Regnault, \textit{Defining a bulk-edge correspondence for non-Hermitian Hamiltonians via singular-value decomposition}, \href{https://journals.aps.org/pra/abstract/10.1103/PhysRevA.99.052118}{Phys. Rev. A \textbf{99}, 052118 (2019)}.

\bibitem{VyasRoy} V.M. Vyas and D. Roy, {\em Topological aspects of periodically driven non-Hermitian Su-Schrieffer-Heeger model}, \href{https://journals.aps.org/prb/abstract/10.1103/PhysRevB.103.075441}{Phys. Rev. B \textbf{103}, 075441 (2021)}.

\bibitem{NehraRoy} R. Nehra and D. Roy, {\em Topology of multipartite non-Hermitian one-dimensional systems}, \href{https://journals.aps.org/prb/abstract/10.1103/PhysRevB.105.195407}{Phys. Rev. B \textbf{105}, 195407 (2022)}.




\bibitem{XiDeWaZhWaYiXu2019} L. Xiao, T. Deng, K. Wang, G. Zhu, Z. Wang, W. Yi, and P. Xue, \textit{Non-Hermitian bulk-boundary correspondence in quantum dynamics}, \href{https://www.nature.com/articles/s41567-020-0836-6}{Nat. Phys. \textbf{16}, 761 (2020)}.


\bibitem{HeHoImAbKiMoLeSzGrTh2019} T. Helbig, T. Hofmann, S. Imhof, M. Abdelghany, T. Kiessling, L.W. Molenkamp, C.H. Lee, A. Szameit, M. Greiter, and R. Thomale, \textit{Generalized bulk-boundary correspondence in non-Hermtitian topolectrical circuits}, \href{https://www.nature.com/articles/s41567-020-0922-9}{Nat. Phys. \textbf{16}, 747 (2020)}.
	
\bibitem{HoHeScSaBrGrKiWoVoKaLeBiThNe2019} T. Hofmann, T. Helbig, F. Schindler, N. Salgo, M. Brzezi\'{n}ska, M. Greiter, T. Kiessling, D. Wolf, A. Vollhardt, A. Kaba\v{s}i, C.H. Lee, A. Bilu\v{s}i\'{c}, R. Thomale, and T. Neupert, \textit{Reciprocal skin effect and its realization in a topolectrical circuit}, \href{https://journals.aps.org/prresearch/abstract/10.1103/PhysRevResearch.2.023265}{Phys. Rev. Res. \textbf{2}, 023265 (2020)}.

\bibitem{GhBrWeCo2019} A. Ghatak, M. Brandenbourger, J. van Wezel, and C. Coulais, \textit{Observation of non-Hermitian topology and its bulk–edge correspondence in an active mechanical metamaterial}, \href{https://www.pnas.org/doi/full/10.1073/pnas.2010580117}{Proc. Natl. Ac. Sc. U.S.A. \textbf{117} 29561 (2020)}.

\bibitem{Okuma2020} N. Okuma, K. Kawabata, K. Shiozaki, and M. Sato, \textit{
Topological Origin of Non-Hermitian Skin Effects}, \href{https://journals.aps.org/prl/abstract/10.1103/PhysRevLett.124.086801}{Phys. Rev. Lett. \textbf{124}, 086801 (2020)}.
 
\bibitem{Okuma2022} N. Okuma and S. Masatoshi, \textit{Non-Hermitian Topological Phenomena: A Review}, \href{https://arxiv.org/abs/2205.10379}{arxiv:2205.10379}.

\bibitem{LinTaiLiLee}
R. Lin, T. Tai, L. Li, and C.H. Lee, \textit{Topological Non-Hermitian skin effect}, \href{https://arxiv.org/abs/2302.03057}{arXiv:2302.03057}.


\bibitem{Longhi2021} S. Longhi, \textit{Non-Hermitian skin effect beyond the tight-binding models}, \href{https://journals.aps.org/prb/abstract/10.1103/PhysRevB.104.125109}{Phys. Rev. B \textbf{104}, 125109 (2021)}.

\bibitem{QinMaShen} F. Qin, Y. Ma, R. Shen, and C.H. Lee, \textit{Universal competitive spectral scaling from the critical non-Hermitian skin effect}. \href{https://arxiv.org/abs/2212.13536}{arXiv:2212.13536}.

\bibitem{WangWang} H. Wang, F. Song, and Z. Wang, \textit{Amoeba formulation of the non-Hermitian skin effect in higher dimensions}, \href{https://arxiv.org/abs/2212.11743}{arXiv:2212.11743}.

\bibitem{JeonLee}
J. Jeon and S. Lee, \textit{Control of localization in non-Hermitian systems}, \href{https://arxiv.org/abs/2211.14336}{arXiv:2211.14336}.

\bibitem{Franca} S. Franca, V. Könye, F. Hassler, J. van den Brink, and C. Fulga, \textit{Non-Hermitian Physics without Gain or Loss: The Skin Effect of Reflected Waves}, \href{https://journals.aps.org/prl/abstract/10.1103/PhysRevLett.129.086601}{Phys. Rev. Lett. \textbf{129}, 086601 (2022)}.

\bibitem{Fan} F. Yang, Q.-D. Jiang, and E.J. Bergholtz, \textit{Liouvillian skin effect in an exactly solvable model}, \href{https://journals.aps.org/prresearch/abstract/10.1103/PhysRevResearch.4.023160}{Phys. Rev. Research {\bf 4}, 023160 (2022)}.


\bibitem{LiangXie} Q. Liang, D. Xie, Z. Dong, H. Li, H. Li, B. Gadway, W. Yi, and B. Yan, \textit{Dynamic Signatures of Non-Hermitian Skin Effect and Topology in Ultracold Atoms}, \href{https://journals.aps.org/prl/abstract/10.1103/PhysRevLett.129.070401}{Phys. Rev. Lett. \textbf{129}, 070401 (2022)}.

\bibitem{Yoshida_MSkin2020} T. Yoshida, T. Mizoguchi, and Y. Hatsugai, \textit{Mirror skin effect and its electric circuit simulation}, \href{https://journals.aps.org/prresearch/abstract/10.1103/PhysRevResearch.2.022062}{Phys. Rev. Res. \textbf{2}, 022062(R) (2020)}.




\bibitem{PoBeKuMoSc2015} C. Poli, M. Bellec, U. Kuhl, F. Mortessagne, and H. Schomerus, \textit{Selective enhancement of topologically induced interface states in a dielectric resonator chain}, \href{https://www.nature.com/articles/ncomms7710}{Nat. Commun. \textbf{6}, 6710 (2015)}.

\bibitem{NHdisc}
X.-Q. Sun, P. Zhu, and T.L. Hughes, {\em Geometric Response and
Disclination-Induced Skin Effects in Non-Hermitian Systems}, \href{https://doi.org/10.1103/PhysRevLett.127.066401} {Phys. Rev. Lett. {\bf 127}, 066401 (2021)}.

\bibitem{NHdisloc}
A. Panigrahi, R. Moessner, and B. Roy, {\em Non-Hermitian dislocation modes: Stability and melting across exceptional points}, \href{https://doi.org/10.1103/PhysRevB.106.L041302} {Phys. Rev. B {\bf 106}, L041302 (2022)}.

\bibitem{NHdisloc2}
F. Schindler and A. Prem, {\em Dislocation non-Hermitian skin effect}, \href{https://journals.aps.org/prb/abstract/10.1103/PhysRevB.104.L161106} {Phys. Rev. B {\bf 104}, L161106 (2021)}.

\bibitem{NHdisloc3}
B.A. Bhargava, I.C. Fulga, J. van den Brink, and A.G. Moghaddam, {\em Non-Hermitian skin effect of dislocations and its topological origin}, \href{https://journals.aps.org/prb/abstract/10.1103/PhysRevB.104.L241402} {Phys. Rev. B {\bf 104}, L241402 (2021)}.


\bibitem{GuGao} Z. Gu, H. Gao, H. Xue, J. Li, Z. Su, and J. Zhu, \textit{Transient non-Hermitian skin effect}, \href{https://www.nature.com/articles/s41467-022-35448-2}{Nat Commun \textbf{13}, 7668 (2022)}. 

\bibitem{LiLeeMuGong} L. Li, C.H. Lee, S. Mu, and J. Gong, \textit{Critical non-Hermitian skin effect},
\href{https://doi.org/10.1038/s41467-020-18917-4}{Nat Commun \textbf{11}, 5491 (2020)}.


\bibitem{elisabet}
E. Edvardsson and E. Ardonne, {\em Sensitivity of non-Hermitian systems}, \href{https://journals.aps.org/prb/abstract/10.1103/PhysRevB.106.115107}{Phys. Rev. B {\bf 106}, 11510 (2022)}.

\bibitem{Trefethen} L.N. Trefethen, {\em Pseudospectra of Linear Operators}, \href{https://epubs.siam.org/doi/abs/10.1137/S0036144595295284}{SIAM Review \textbf{39}, Iss. 3 (1997)}.


\bibitem{NHsensor} J.C. Budich and E.J. Bergholtz, \textit{Non-Hermitian topological sensors}, \href{https://journals.aps.org/prl/abstract/10.1103/PhysRevLett.125.180403}{Phys. Rev. Lett. \textbf{125}, 180403 (2020)}.

\bibitem{NHsensor2} F. Koch and J.C. Budich \textit{Quantum non-Hermitian topological sensors}, \href{https://journals.aps.org/prresearch/abstract/10.1103/PhysRevResearch.4.013113}{Phys. Rev. Research \textbf{4}, 013113 (2022)}.

\bibitem{NHsensor3} A. McDonald and A.A. Clerk, \textit{Exponentially-enhanced quantum sensing with non-Hermitian lattice dynamics}, \href{https://www.nature.com/articles/s41467-020-19090-4}{Nat. Commun. \textbf{11}, 5382 (2020)}.

\bibitem{NHsensor4} J. Wiersig, \textit{Review of exceptional point-based sensors}, \href{https://doi.org/10.1364/PRJ.396115}{Photon. Res. \textbf{8}, 1457-1467 (2020)}.

\bibitem{NHsensor5} H. Schomerus, \textit{Nonreciprocal response theory of non-Hermitian mechanical metamaterials: Response phase transition from the skin effect of zero modes}, \href{https://journals.aps.org/prresearch/abstract/10.1103/PhysRevResearch.2.013058}{Phys. Rev. Research \textbf{2}, 013058 (2020)}.


\bibitem{Anderson} P.W. Anderson, \textit{Absence of Diffusion in Certain Random Lattices}, \href{https://journals.aps.org/pr/abstract/10.1103/PhysRev.109.1492}{Phys. Rev. \textbf{109}, 1492 (1958)}.

\bibitem{Thouless} D.J. Thouless, \textit{Electrons in disordered systems and the theory of localization}, \href{https://doi.org/10.1016/0370-1573(74)90029-5}{Phys. Rep. \textbf{13}, 93 (1974)}.

\bibitem{TomKhay} G. De Tomasi and I.M. Khaymovich, \textit{Non-Hermiticity induces localization: good and bad resonances in power-law random banded matrices}, \href{https://arxiv.org/abs/2302.00015}{arXiv:2302.00015}.

\bibitem{SunLiu2023} X.Q. Sun and C.S. Liu, \textit{Localization and topological transitions in non-Hermitian SSH models}, \href{https://arxiv.org/abs/2212.12288v2}{arXiv:2212.12288v2}.

\bibitem{WangWang2023} C. Wang and X.R. Wang, \textit{Anderson localization transitions in disordered non-Hermitian systems with exceptional points}, \href{https://arxiv.org/abs/2209.07072v2}{arXiv:2209.07072v2}.

\bibitem{LuOhtShin} X. Luo, T. Ohtsuki, and R. Shindou, \textit{Transfer matrix study of the Anderson transition in non-Hermitian systems}, \href{https://journals.aps.org/prb/abstract/10.1103/PhysRevB.104.104203}{Phys. Rev. B \textbf{104}, 104203 (2021)}.

\bibitem{ZengChen} Q.B. Zeng, S. Chen, and R. Lü, \textit{Anderson localization in the non-Hermitian Aubry-André-Harper model with physical gain and loss}, \href{https://journals.aps.org/pra/abstract/10.1103/PhysRevA.95.062118}{Phys. Rev. A \textbf{95}, 062118 (2017)}.

\bibitem{SpringKonye} H. Spring, V. Könye, F.A. Gerritsma, I.C. Fulga, and A.R. Akhmerov, \textit{Phase transitions of wave packet dynamics in disordered non-Hermitian systems}, \href{https://arxiv.org/abs/2301.07370v1}{ arXiv:2301.07370v1}.

\bibitem{Hui} H. Liu, J.-S. You, S. Ryu, and I.C. Fulga, \textit{Supermetal-insulator transition in a non-Hermitian network model}, \href{https://journals.aps.org/prb/abstract/10.1103/PhysRevB.104.155412}{ Phys. Rev. B {\bf 104}, 155412 (2021)}.

\bibitem{Hebert} F. Hébert, M. Schram, R.T. Scalettar, W.B. Chen, and Z. Bai, \textit{Hatano-Nelson model with a periodic potential}, \href{https://link.springer.com/article/10.1140/epjb/e2011-10875-9}{Eur. Phys. J. B \textbf{79}, 465–471 (2011)}.

\bibitem{Jiang:2019} 
H. Jiang, L.J. Lang, C. Yang., S.L. Zhu, and S. Chen, \textit{Interplay of non-hermitian skin effects and Anderson localization in nonreciprocal quasiperiodic lattices}, \href{DOI:https://doi.org/10.1103/PhysRevB.100.054301}{Phys. Rev. B \textbf{100}, 054301 (2019).}

\bibitem{LiuZhouChen2021} Y. Liu, Q. Zhou, and S. Chen, \textit{Localization transition, spectrum structure, and winding numbers for one-dimensional non-Hermitian quasicrystals}, \href{https://journals.aps.org/prb/abstract/10.1103/PhysRevB.104.024201}{Phys. Rev. B \textbf{104}, 024201 (2021)}.

\bibitem{Longhi2019} S. Longhi, \textit{Non-Hermitian topological phase transition in PT-symmetric mode-locked lasers}, \href{https://opg.optica.org/ol/fulltext.cfm?uri=ol-44-5-1190&id=405932}{Opt. Lett. \textbf{44}, 1190-1193 (2019)}.

\bibitem{Longui20192} S. Longhi, \textit{Metal-insulator phase transition in a non-Hermitian Aubry-André-Harper model}, \href{https://journals.aps.org/prb/abstract/10.1103/PhysRevB.100.125157}{Phys. Rev. B \textbf{100}, 125157 (2019)}.

\bibitem{OrImu} T. Orito and K.I. Imura, \textit{Unusual wave-packet spreading and entanglement dynamics in non-Hermitian disordered many-body systems}, \href{https://journals.aps.org/prb/abstract/10.1103/PhysRevB.105.024303}{Phys. Rev. B \textbf{105}, 024303 (2022)}.

\bibitem{ChenChengLin} W. Chen, S. Cheng, J. Lin, R. Asgari, and G. Xianlong, \textit{Breakdown of the correspondence between the real-complex and delocalization-localization transitions in non-Hermitian quasicrystals}, \href{https://journals.aps.org/prb/abstract/10.1103/PhysRevB.106.144208}{Phys. Rev. B \textbf{106}, 144208 (2022)}.

\bibitem{Longui2021} S. Longhi, \textit{Phase transitions in a non-Hermitian Aubry-André-Harper model}, \href{https://journals.aps.org/prb/abstract/10.1103/PhysRevB.103.054203}{Phys. Rev. B \textbf{103}, 054203 (2021)}.

\bibitem{LinLiXiaoWangYiXue} Q. Lin, T. Li, L. Xiao, K. Wang, W. Yi, and P. Xue, {\em Observation of non-Hermitian topological Anderson insulator in quantum dynamics}, \href{https://www.nature.com/articles/s41467-022-30938-9}{Nat Commun {\bf 13}, 3229 (2022)}.

\bibitem{LinLiXiaoWangYiXuePRL} Q. Lin, T. Li, L. Xiao, K. Wang, W. Yi, and P. Xue, {\em Topological Phase Transitions and Mobility Edges in Non-Hermitian Quasicrystals}, \href{https://journals.aps.org/prl/abstract/10.1103/PhysRevLett.129.113601}{Phys. Rev. Lett. {\bf 129}, 113601 (2022)}.



\bibitem{Stocker:2022}
L. Stocker, S.H. Sack, M.S. Ferguson, and O. Zilberberg, \textit{Entanglement based observables for quantum impurities}, \href{https://doi.org/10.1103/PhysRevResearch.4.043177}{Phys. Rev. Research \textbf{4}, 043177 (2022)}.

\bibitem{Capizzi:2023}
L. Capizzi, S. Scopa, F. Rottoli, and P. Calabrese, \textit{Domain wall melting across a defect}, \href{https://doi.org/10.1209/0295-5075/acb50a}{Europhys. Lett. \textbf{141}, 31002 (2023)}.

\bibitem{Brauneis:2023}
F. Brauneis, A. Ghazaryan, H.-W. Hammer, and A.G. Volosniev, \textit{Emergence of a Bose polaron in a small ring threaded by the Aharonov-Bohm flux}, \href{https://doi.org/10.48550/arXiv.2301.10488}{arxiv:2301.10488 (2023)}.

\bibitem{Andreanov:2023}
A. Andreanov, M. Carrega, J. Murugan, J. Olle, D. Rosa, and R. Shir, \textit{From Dyson Models to Many-Body Quantum Chaos}, \href{https://doi.org/10.48550/arXiv.2302.00917}{arxiv:2302.00917 (2023)}.

\bibitem{ShenChenQinZhongLee} R. Shen, T. Chen, F. Qin, Y. Zhong, and C.H. Lee, {\em Proposal for observing Yang-Lee criticality in Rydberg atomic arrays}, \href{https://arxiv.org/abs/2302.06662}{arXiv:2302.06662v1}.

\bibitem{QinShenLee2023} F. Qin, R. Shen, and C.H. Lee, {\em Non-Hermitian squeezed polarons}, \href{https://journals.aps.org/pra/abstract/10.1103/PhysRevA.107.L010202}{Phys. Rev. A \textbf{107}, L010202 (2023)}.



\bibitem{scalefree21}
L. Li, C.H. Lee, and J. Gong, {\em Impurity induced scale-free localization}, \href{https://www.nature.com/articles/s42005-021-00547-x} {Communications Physics {\bf 4}, 42 (2021)}.


\bibitem{HatanoNelson} N. Hatano and D.R. Nelson, \textit{Localization Transitions in Non-Hermitian Quantum Mechanics}, \href{https://journals.aps.org/prl/abstract/10.1103/PhysRevLett.77.570}{Phys. Rev. Lett. \textbf{77}, 570 (1996)}.

\bibitem{HatanoNelson2} N. Hatano and D.R. Nelson, \textit{Vortex pinning and non-Hermitian quantum mechanics}, \href{https://journals.aps.org/prb/abstract/10.1103/PhysRevB.56.8651}{Phys. Rev. B \textbf{56}, 8651 (1997)}.

\bibitem{GuoLiuZhaoLiuChe2021} C.-X. Guo, C.-H. Liu, X.-M. Zhao, Y. Liu, and S. Chen, \textit{Exact Solution of Non-Hermitian Systems with Generalized Boundary Conditions: Size-Dependent Boundary Effect and Fragility of the Skin Effect}, \href{https://journals.aps.org/prl/abstract/10.1103/PhysRevLett.127.116801}{Phys. Rev. Lett. \textbf{127}, 116801 (2021)}.


\bibitem{HatanoNelson3} N. Hatano and D.R. Nelson, \textit{Non-Hermitian delocalization and eigenfunctions}, \href{https://journals.aps.org/prb/abstract/10.1103/PhysRevB.58.8384}{Phys. Rev. B \textbf{ 58}, 8384 (1998)}.



\bibitem{note} A similar argument holds for the \emph{left} eigenvectors, which are localized on the opposite end.
%

\bibitem{SuGuWaLiRuDuCheZhe} L. Su, C.-X. Guo, Y. Wang, L. Li, X. Ruan, Y. Du, S. Chen, and D. Zheng, \textit{Observation of size-dependent boundary effects in non-Hermitian electric circuits}, \href{http://cpb.iphy.ac.cn/EN/abstract/abstract125515.shtml}{Chin. Phys. B \textbf{32}, 038401 (2023)}.


\bibitem{supmat} See supplementary material.


\bibitem{scalefree23}
C.-X. Guo, X. Wang, H. Hu, and S. Chen, {\em Accumulation of scale-free localized states induced by local non-Hermiticity}, \href{https://arxiv.org/abs/2302.02798} {arXiv:2302.02798}.


\bibitem{scalefree232}
B. Li, H.-R. Wang, F. Song, and Z. Wang, {\em Scale-free localization and PT symmetry breaking from local non-Hermiticity}, \href{https://arxiv.org/abs/2302.04256} {arXiv:2302.04256}.


\bibitem{impurityproblem2}
Y. Liu, Y. Zeng, L. Li, and S. Chen, {\em Exact solution of the single impurity problem in nonreciprocal lattices: Impurity-induced size-dependent non-Hermitian skin effect}, \href{https://journals.aps.org/prb/abstract/10.1103/PhysRevB.104.085401} {Phys. Rev. B {\bf 104}, 085401 (2021)}.

\bibitem{impurityproblem}
F. Roccati, {\em Non-Hermitian skin effect as an impurity problem}, \href{https://journals.aps.org/pra/abstract/10.1103/PhysRevA.104.022215} {Phys. Rev. A {\bf 104}, 022215 (2021)}.


\bibitem{balatsky}
P.O. Sukhachov and A.V. Balatsky, {\em Non-Hermitian impurities in Dirac systems}, \href{https://journals.aps.org/prresearch/abstract/10.1103/PhysRevResearch.2.013325}{Phys. Rev. Research {\bf 2}, 013325 (2020)}.

\bibitem{StegImHelHofLee} 
A. Stegmaier, S. Imhof, T. Helbig, T.Hofmann, C.H. Lee, M. Kremer, A. Fritzsche, T. Feichtner, S. Klembt, S. Höfling, I. Boettcher, I.C. Fulga, L. Ma, O.G. Schmidt, M. Greiter, T. Kiessling, A. Szameit, and R. Thomale, \textit{Topological Defect Engineering and PT Symmetry in Non-Hermitian Electrical Circuits}, \href{https://journals.aps.org/prl/abstract/10.1103/PhysRevLett.126.215302}{Phys. Rev. Lett. \textbf{126}, 215302 (2021)}.




\bibitem{ChalkerMehlig} J.T. Chalker and B. Mehlig, \textit{Eigenvector Statistics in Non-Hermitian Random Matrix Ensembles}, \href{https://journals.aps.org/prl/abstract/10.1103/PhysRevLett.81.3367}{Phys. Rev. Lett. \textbf{81}, 3367 (1998)}.


\bibitem{funnel} S. Weidemann, M. Kremer, T. Helbig, T. Hofmann, A. Stegmaier, M. Greiter, R. Thomale, and A. Szameit, \textit{Topological funneling of light}, \href{https://www.science.org/doi/full/10.1126/science.aaz8727}{Science \textbf{368}, 311 (2020)}.

 \bibitem{photonics}
 T. Ozawa, H.M. Price, A. Amo, N. Goldman, M. Hafezi, L. Lu, M.C. Rechtsman, D. Schuster, J. Simon, O. Zilberberg, and I. Carusotto, {\em Topological Photonics}, \href{https://journals.aps.org/rmp/abstract/10.1103/RevModPhys.91.015006}{Rev. Mod. Phys. {\bf 91}, 015006 (2019)}.

\bibitem{bethescalefree} H.-R. Wang, B. Li, F. Song, and Z. Wang, \textit{Scale-free non-Hermitian skin effect in a boundary-dissipated spin chain}, \href{https://arxiv.org/abs/2301.11896}{arXiv:2301.11896}.


\bibitem{Xiao:2022} Y.-X. Xiao and C.T. Chan, \textit{Topology in non-Hermitian Chern insulators with skin effect}, Phys. Rev. B \textbf{105}, 075128 (2022).

\end{thebibliography}
\end{document}